# Optical bistable SOI micro-ring resonators for memory applications

Andrey A. Nikitin[1], Ilya A. Ryabcev[1], Aleksei A. Nikitin[1], Alexander V. Kondrashov[1], Alexander A. Semenov[1], Dmitry A. Konkin[2,3], Andrey A. Kokolov[2,4], Feodor I. Sheyerman[2], Leonid I. Babak[2], and Alexey B. Ustinov[1]

[1] *St. Petersburg Electrotechnical University "LETI", St. Petersburg, Russia*

[2] *Tomsk State University of Control Systems and Radioelectronics "TUSUR", Tomsk, Russia*

[3] *National Research Tomsk Polytechnic University, Tomsk, Russia*

[4] *V.E. Zuev Institute of Atmospheric Optics SB RAS, Tomsk, Russia*

The present work focuses on experimental investigations of a bistabile silicon-on-insulator (SOI) micro-ring resonator (MRR). The resonator exploits a continuous-wave operation of the carrier-induced bistability demonstrating a stable hysteresis response at the through and drop ports when the input power exceeds the threshold value. Flipping the optical input power provides a convenient mechanism for a switching of the MRR output characteristics between two steady states having a long holding time. The transition of the resonator output between these states is experimentally investigated. It is shown that the switching speed is limited by a low-to-high transition of 188 ns. Obtained results shows an application of the passive SOI MRR as an all-optical memory cell with two complementary outputs.

## I. INTRODUCTION

In the last decade, the progress in the development of photonic integrated circuits has opened new avenues for an optical memory [1]. The most common approaches for a light storage rely on a bistability phenomenon. The bistability of optical systems allows switching between its two stable states under external conditions. One way to provide the bistable behavior is to utilize a dispersive nonlinearity in the photonic crystals [2–7] that demonstrate a clear hysteresis response. The bistable behavior of these structures manifests itself due to the carrier generation induced by two-photon absorption (TPA). This effect causes the free-carrier dispersion (FCD) resulting in a reduction of the refractive index and yields an upward frequency shift of resonant harmonics providing a hysteresis response. Nevertheless, the holding time is limited by a heating due to an opposite action of the thermo-optic effect suppressing the bistability induced by the carrier effect. Laser-based bistable devices address this issue and provide a long holding time, since lasing continues even with the thermally induced frequency shift [1,8,9].

Another approach for the optical memory implementation is to exploit the feedback loop with an active nonlinear element [10–15]. A replacement of a lumped nonlinear active element with distributed one provides an elegant solution for miniaturization and further improvement of the memory performance. Ring and microdisk lasers are fascinating examples that obtain a highly compact integrated memory schemes with a low switching power [16,17]. These schemes require either an optical pumping at a different wavelength from the signal or an electrical pumping [18]. At the same time, the dispersive nonlinearity of the passive micro-ring resonators (MRRs) attracts research interest due to manifold nonlinear dynamics promising for a wide variety of applications [19,20]. However, technological constraints inherent to optical memories based on passive semiconductor MRRs are related to the dominant role of a slow thermo-optic effect over fast nonlinear ones. Owing to this feature, the functionality of all-optical memories based on passive MRRs reported in the previous studies relies on the thermal nonlinear effect [21,22]. In these works, the switching time is as low as 1 µs and 253 ns for the silicon and graphene-on-silicon nitride chips, respectively. These transient processes are governed by the thermal dissipation in the cavities.

The Kerr or free-carrier effects pave the way to the sufficient improvement of the switching time of the nonlinear resonator. Both of them can be considered instantaneous in comparison with the thermal nonlinear effect [18,23,24]. The asymmetric Kerr-induced bistability between counter propagating light states opens new operating regimes for a realization of bi-directional all-optical memories and switches that was recently demonstrated for the fused silica micro-resonator [18]. However, this cavity was fabricated from a 3-mm-diameter silica rod, which is not compatible with the integrated circuit technology. The FCD-induced optical bistability effect in a silicon ring resonator with a nanosecond transition time was experimentally demonstrated in [25]. This is fascinating effect for a realization of high-speed all-optical on-chip memories. However, the interplay between the TPA and temperature drift hinders the long-term stability in silicon MRRs [25–29]. This interplay produces a contradiction between a long storage time and high switching speed.

A perspective approach for all-optical memories is to use the MRR fabricated with silicon-on-insulator (SOI) technology demonstrating the FCD-induced bistability under a continuous-wave pumping due to its design [30]. This design naturally circumvents the abovementioned contradiction and offers the significant breakthrough towards passive all-optical memories compatible with CMOS SOI processes.

Purpose of the present work is to study the SOI MRR response for the pulsed optical pumping. Switching time between two stable states with high and low absorption is investigated for different optical pulse width and intensity. As a result, we show, for the first time, a robust operation of the SOI micro-ring resonator as an all-optical memory cell driven by the FCD-induced bistability paving the way for fabrication of monolithically-integrated optical random access memories and optical memories.

## II. EXPERIMENTAL SETUP

A symmetrically coupled micro-ring resonator is fabricated on a silicon-on-insulator substrate with a 2-µm-thick buried oxide layer by the IHP's SG25PIC technology. The thickness of a cladding $SiO_2$ layer is 14 µm. A width and height of silicon waveguides are 500 nm and 220 nm, respectively. A diameter of the ring is 256 µm. A distance between the MRR and the straight input and output waveguides is 250 nm.

The experimental setup is shown in Fig. 1. A light from the 1550-nm tunable laser is applied to an optical isolator (OI) preventing the backward propagation of the light. The tunable laser allows to measure the MRR frequency response by a continuous frequency sweep in both upwards and downwards directions. An erbium-doped fiber amplifier (EDFA) is used to control the power at the input port of the resonator. A two-port arbitrary waveform generator (AWG) and Mach-Zehnder electro-optic modulator (EOM) serve for the fast switching of the MRR between the stable states. A polarization controller adjusts the light polarization to enhance both the coupling efficiency and output power. The light is coupled through the out-of-plane fibers to the Bragg grating couplers providing the linear fiber-to-fiber insertion loss of 12.5 dB. Output signals at the through and drop ports of the MRR are split with ratio of 4:1 into two channels. The first channel is connected to an optical spectrum analyzer (OSA) in order to obtain the resonator power-frequency response. This response is recorded as a maximum power of the transmitted optical harmonic having a slow variation in frequency. This harmonic is generated initially with the single-frequency tunable laser. The second channel is connected to a photodetector and oscilloscope to measure the temporal response of the resonator.

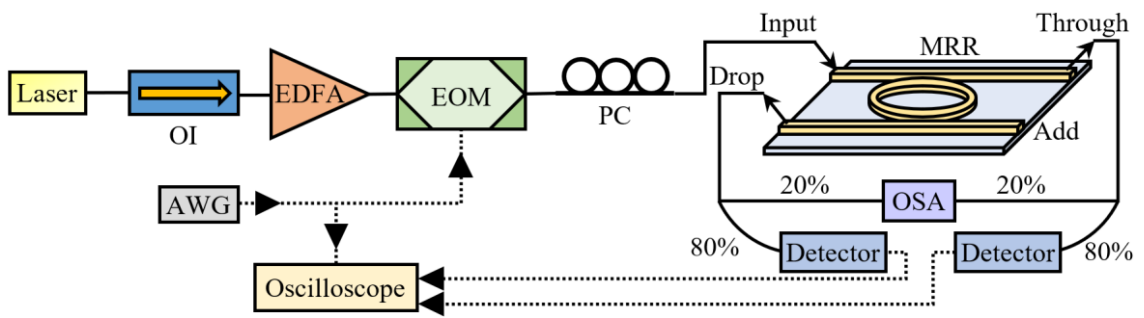

Fig. 1. Schematic representation of the experimental setup. Solid lines show optical fibers. Dotted lines show microwave cables.

## III. BISTABLE BEHAVIOR OF SOI MICRO-RING RESONATOR

In this Section, the bistable behavior of the SOI micro-ring resonator shown in Fig. 1 is investigated in a two-step procedure. As a first step, the resonator frequency response is measured with up- and down-sweeping frequency of the light with the rate 1 GHz/s. The micro-ring resonator demonstrates multiresonant transmission characteristics with the distance between the neighboring frequencies, i.e. the free spectral range, of 88.8 GHz. Fig. 2 (a) shows the fragments of these characteristics measured at the through and drop ports by black solid and dashed lines, respectively. To ensure the linear operation, low optical input power of -10 dBm is applied to the MRR input. The loaded Q-factor for this input power is found to be 45000. The measurement is carried out in the vicinity of the resonant harmonic of 191.811 THz. Further, we restrict ourselves to the investigation of a bistable behavior of this resonant harmonic.

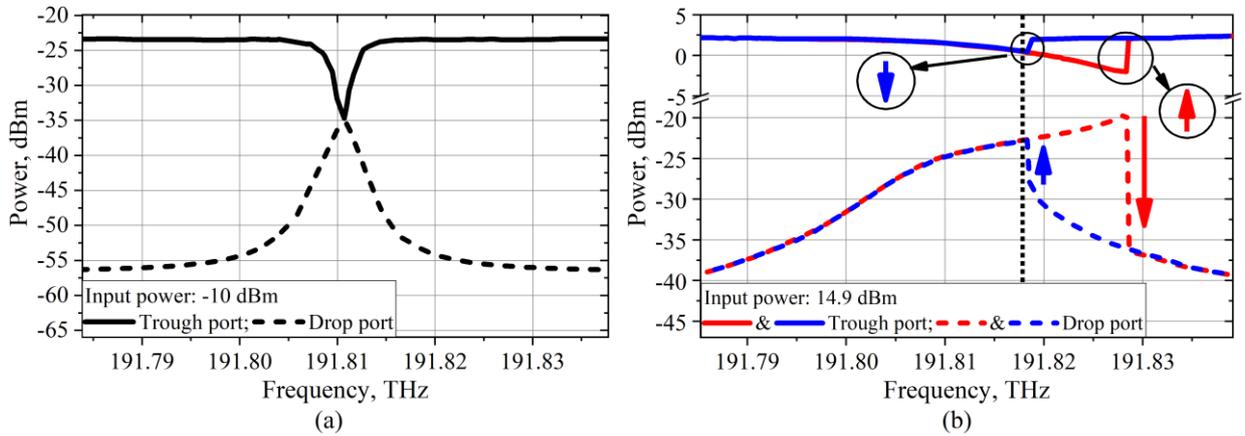

Fig. 2. Linear frequency responses of the MRR at the through (solid line) and drop (dashed line) ports (a); nonlinear frequency responses of the MRR at the through (solid lines) and drop (dashed lines) ports measured for the upwards (red lines) and downwards (blue lines) frequency sweep (b).

Owing to the carrier effect, an increase of the input power leads to a positive shift of the resonant frequency. As soon as the input power exceeds the bistability threshold value of 3.6 dBm, the hysteresis loop appears at the frequency response. Fig. 2(b) shows the nonlinear frequency response obtained for the input power of 14.9 dBm. Here, solid and dashed lines represent characteristics taken at the through and drop ports of the resonator, respectively. Red and blue lines in Fig. 2(b) indicate the upwards and downwards frequency sweep, respectively. As is seen, both dependencies are characterized by the bistable hysteresis loop with the width of 10 GHz. Note that observed bistable hysteresis response is driven by the free-carrier dispersion effect, which dominates the thermal nonlinearity [30]. The frequency response at the drop port demonstrates the high-to-low and low-to-high transitions as the frequency is swept upwards and downwards, respectively, while the through port shows the opposite behavior. This feature of the frequency response is of a considerable importance for memory and logic applications, since the output states at the through and drop ports are complementary.

As a second step, we investigate the transition time between low and high states of the MRR output. The experiment is carried out as follows. The frequency of the input light is chosen to be 191.818 THz that is shown by black dotted line in Fig. 2(b). A slow variation of the input power from -10 dBm to 14.9 dBm provides fast switches of the output signal shown by arrows in Fig. 2(b). In order to implement this slow variation, a triangular electric signal with a frequency of 100 Hz is applied to the input port of the EOM. The dependence of the input power versus time is shown by gray lines in Fig. 3(a) and 3(b). Here, all input and output characteristics are normalized to the maximum power.

Output waveforms taken from the through and drop ports are shown by blue and red lines in Fig. 3(a) and 3(b), respectively. Let us consider obtained characteristics. The input light power is increased during the first half of the period (see gray lines in Fig. 3(a) and 3(b)). At the very beginning, it provides a rise of the output power at both ports. A sharp switch of the output power in opposite directions is observed at 3.9 ms. While the output power at the through port is fall into the low state, the power at the drop port is raised to the high state. After 5 ms the input power decreases during the second half of the period. The transitions into opposite states are observed at 8.6 ms. During experiments, a special attention is given to the investigation of the transition times of the output characteristic shown in Fig. 3(a). The investigation is carried out for various input power levels. Fig. 3(c) shows the set (low-to-high transition) and reset (high-to-low transition) times as a function of the input power. It is seen that the increase in the input power reduces the transition times that reach the minima of 28 ns and 188 ns for reset and set times, respectively.

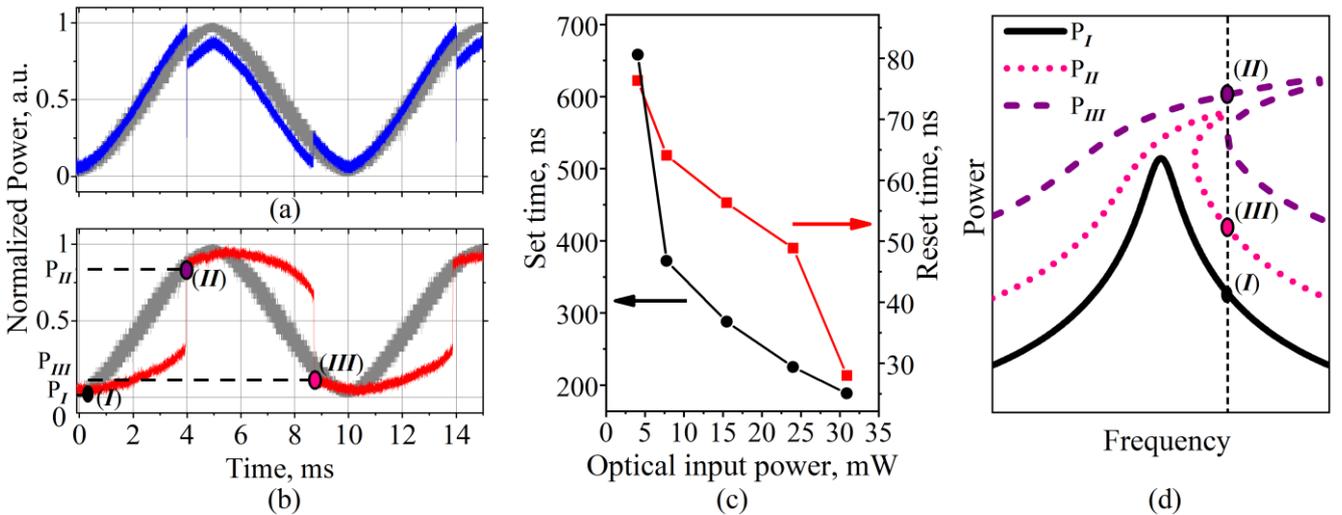

Fig. 3. Input and output waveforms taken from the through (a) and drop (b) ports versus time; dependencies of the set and reset time versus the input power (c); qualitative dependencies of the power at the drop port versus frequency (d).

The behavior observed in Fig. 3(a) and 3(b) can be explained using the qualitative dependencies of the output power at the drop port of the MRR versus frequency for the various input powers shown in Fig. 3(d). These powers correspond to the time points I, II, and III shown in Fig. 3(b). Black dashed line in Fig. 3(d) demonstrates the operating frequency of the light. Initially, the resonator operates linearly while the input power is low (see point I in Fig. 3(b)). In this case, the output power corresponds to the low state (point I in Fig. 3(d)). An increase of the input intensity provides a FCD-induced upward frequency shift [25,30]. Further intensity increase above the bistability threshold value leads to an extension of the bistability frequency range, as well as upwards shift of this region [31]. At the time point II (see Fig. 3(b)), the bottom frequency border of the bistability range shifts beyond the frequency of the operating signal. In this case, only the single stable state with a high output power is available. This corresponds to switching of the MRR to a high output state (see point II in Fig. 3(d)). The resonator holds this state while the upper frequency border of the bistability range is above the operating frequency. A decrease in the input power provides a downwards shift of the bistability borders [30]. At the time point III shown in Fig. 3(b), the upper frequency border drops below the operating frequency, therefore the MRR switches back into the low state (see point III in Fig. 3(d)).

The dependencies of the output versus input powers are extracted for both ports using the time synchronization of the characteristics shown in Fig. 3(a) and 3(b). As is seen in Fig. 4, the obtained dependencies are characterized by the power hysteresis loops. Arrows in Fig. 4 show the motion directions along the hysteresis loop when the input power is varied. The sufficient change in the input power causes the switching of the output power. Such behavior represents a key effect for memory applications. Let us exemplify this claim. In order to accomplish this, the operating power at the level of 0.5 a.u. is chosen. This level is marked by dashed line in Fig. 4(a) and 4(b). Assume that the output signal at the drop port is conformed to the low state marked by open red circle in Fig. 4(a). This scenario corresponds to the high state at the through port as well (see open red circle in Fig. 4(b)). In order to switch the states of the output signal, the short-term increase of the input power above 0.8 a.u. is required. In this case, the return of the input signal to the operating power level of 0.5 a.u. is occurred at the second branch of the hysteresis loop. It means a switch of the output power into the opposite states shown by open blue circles in Fig. 4(a) and 4(b). The switch back into the initial states is realized due to a short-term decrease of the input power below 0.15 a.u. Obtained low and high output states are regarded as logical "0" and "1", respectively.

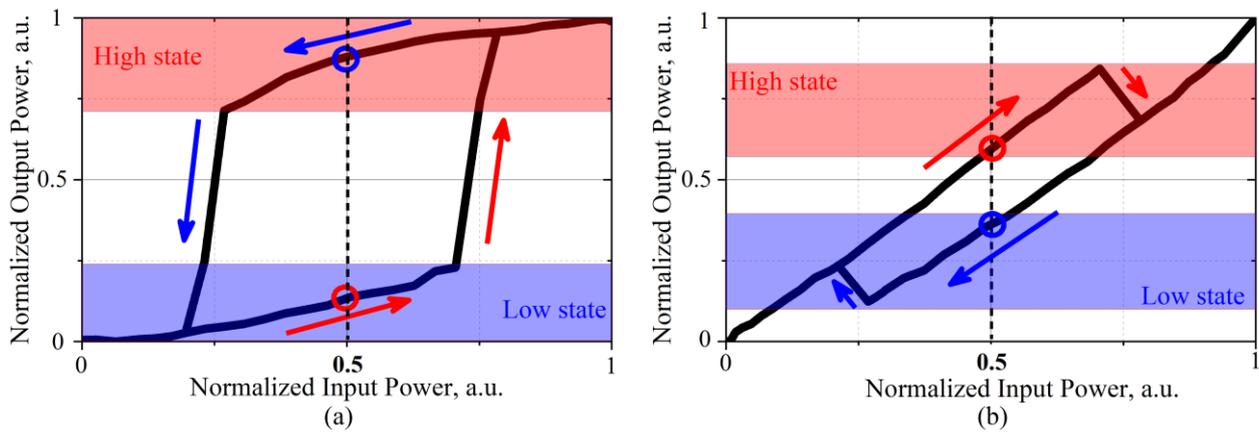

Fig. 4. Dependencies of the output power at the drop (a) and through (b) ports versus the input power.

## VI. OPERATION OF THE SOI MICRO-RING RESONATOR AS A MEMORY CELL

This Section demonstrates an operation of the MRR as a memory cell. The bistable behavior of the silicon ring resonator was discussed in previous Section. It was shown that a switching of the MRR output is occurred when the optical input power flips between two particular levels. Consider now operation of the SOI MRR as a memory cell. In order to accomplish this, the electric signal with the frequency of 100 kHz is applied to the EOM, which modulates the light intensity. The frequency of the light is 191.818 THz. The input optical signal detected by the photodetector is shown in Fig. 5(a). The lengths of the positive and negative pulses are equal to 700 ns. The resonator outputs are detected by the photodetector and recorded by the oscilloscope. Fig. 5(b) and 5(c) show the experimentally observed temporal responses of the MRR measured at the drop and through ports, respectively. As is seen, the positive pulse results in the switch of the drop port output into the high state ("1") while the through port output transits into the low state ("0"). The opposite transitions of the output characteristics are observed when the negative pulses are appeared at the input. Thus the through and drop ports are complementary to each other. As is seen in Fig. 5(b) and 5(c), the holding time for applied input signal is about 4 μs. Additional experiments demonstrate the long-term stability of the output power for more than one hour. The minimum pulse length, which permits the stable memory operation, is measured to be 200 ns. This pulse length provides the maximum set-reset frequency of 2.5 MHz. These limits are defined by the set time of 188 ns measured in previous Section.

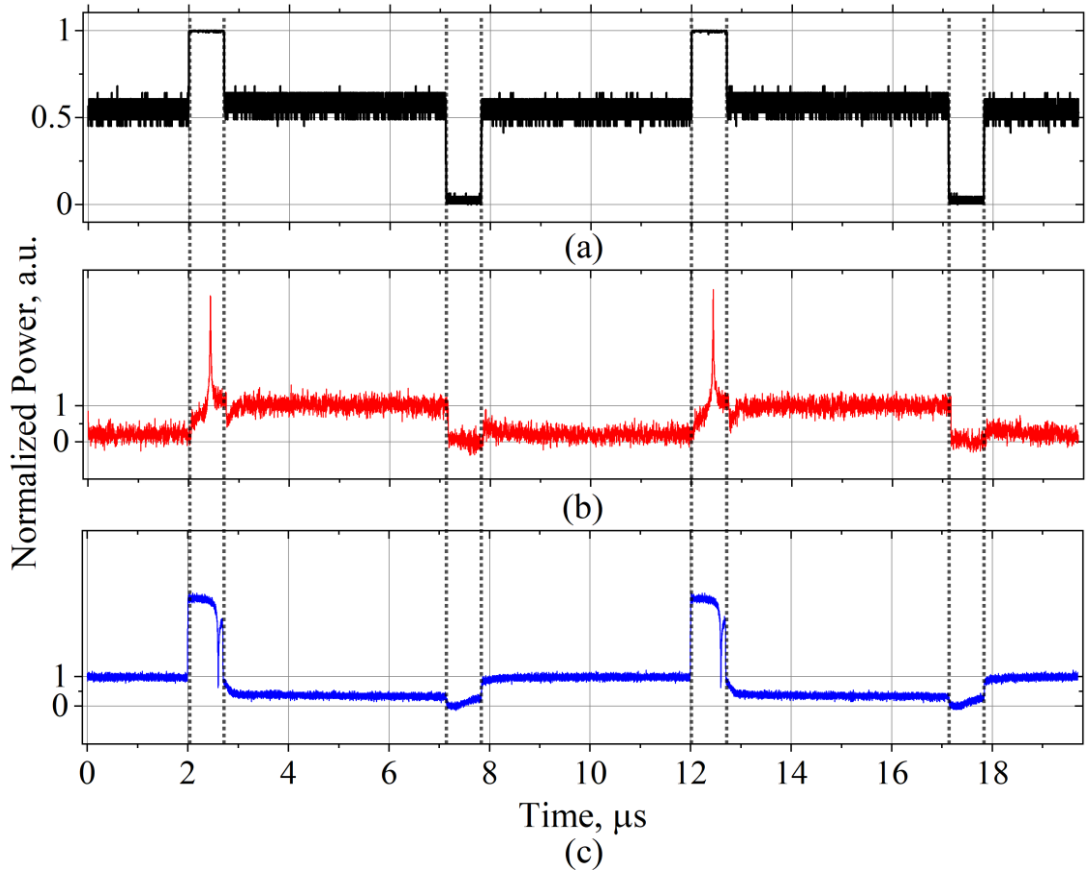

Fig. 5. Input optical signal (a); temporal responses of the micro-ring resonator from the drop (b) and through (c) ports.

## VII. CONCLUSION

Current work demonstrates a proof-of-concept of the all-optical memory cell based on the passive SOI micro-ring resonator. The memory cell utilizes the bistable behavior driven by a TPA-induced free-carrier dispersion. It is shown that MRR outputs transit between two stable states due to the variation of the input power when the operating frequency is closed to the bistable frequency range. The transition time between low and high states of the MRR output is investigated. The set time of 188 ns and reset time of 28 ns are measured at the through port for input power level of 14.9 dBm. The MRR holds the output state for more than one hour. The output characteristics measured at the through and drop ports of the MRR manifest a complimentary behavior. While the positive pulse sets the output of the drop port into the high state, the through port is simultaneously reset it into the low state. In the meantime, the negative pulse applied to the ring produces opposite transitions at the output ports. The operating frequency is limited by 2.5 MHz mainly due to a quite slow set time. This limits the switching speed between two output states and provides a technological constraint inherent to the proposed memory technique. In silicon waveguides, this speed is defined by an effective carrier recombination time that depends on various factors as an employed SOI fabrication process and typically does not exceed 200 ns [32,33]. Another restriction of the switching speed issues from the quality factor of the MRR. The resonator with a higher quality factor requires more time to transit between its states. Both factors can be sufficiently improved by an optimization of the resonator geometry, by utilizing the ion implantation, and by the free carrier diffusion [24,32,33]. These methods provide elegant ways to push the limits of the SOI technology. Therefore, the considered passive MRR looks favorable for not only investigations of new physical phenomena but also for applications as a complimentary part to the traditional approach for general computing and signal processing.

## ACKNOWLEDGMENTS

The work of St. Petersburg Electrotechnical University was supported by the Ministry of Science and Higher Education of the Russian Federation (grant number No. FSEE-2020-0005). The work of Tomsk State University of Control Systems and Radioelectronics was supported by the Ministry of Science and Higher Education of the Russian Federation (unique identifier is FEWM-2020-0046) and RSF grant № 21-79-10077. The work of V.E. Zuev Institute of Atmospheric Optics SB RAS was supported by the Ministry of Science and Higher Education of the Russian Federation (unique identifier is AAAA-A19-119110690036-9).